\documentclass[conference]{IEEEtran}
\IEEEoverridecommandlockouts
\usepackage{cite}
\usepackage{amsmath,amssymb,amsfonts}
\usepackage{amsthm}
\usepackage{algorithmic}
\usepackage{graphicx}
\usepackage{textcomp}
\usepackage{xcolor}
\usepackage{lipsum}  
\usepackage{mathtools}
\usepackage{todonotes}
\usepackage{subfig}
\usepackage{siunitx}
\def\BibTeX{{\rm B\kern-.05em{\sc i\kern-.025em b}\kern-.08em
    T\kern-.1667em\lower.7ex\hbox{E}\kern-.125emX}}
\DeclarePairedDelimiter{\floor}{\lfloor}{\rfloor}
\DeclarePairedDelimiter{\ceil}{\lceil}{\rceil}

\usepackage{url}

\begin{document}
\title{Ensuring Reliable and Predictable Behavior of 
IEEE 802.1CB Frame Replication and Elimination}

\author{\IEEEauthorblockN{Lisa Maile, Dominik Voitlein, Kai-Steffen Hielscher, Reinhard German}
\IEEEauthorblockA{Computer Networks and Communication Systems\\
Friedrich-Alexander University Erlangen-Nürnberg, Germany\\
\{lisa.maile, dominik.voitlein, kai-steffen.hielscher, reinhard.german\}@fau.de}
}

\maketitle

\begin{abstract}
Ultra-reliable and low-latency communication has received significant research attention. A key part of this evolution are the Time-Sensitive Networking (TSN) standards, which extend Ethernet with real-time mechanisms. To guarantee high reliability, the standard IEEE 802.1CB-2017 \textit{Frame Replication and Elimination for Reliability }enables redundant communication over disjoint paths. While this mechanism is essential for time-critical applications, the standard contains some fundamental limitations that can compromise safety. Although some of these limitations have been addressed, none of the previous works provide solutions to these problems. This paper presents solutions to four main limitations of the IEEE 802.1CB-2017 standard. These are 1) choosing match versus vector recovery algorithm, 2) defining the length of the sequence history, 3) setting a timer to reset the sequence history, and 4) dimensioning the burst size in case of link failures. We show how these challenges can be solved by using best- and worst-case path delays of the network. We have performed simulations to illustrate the impact of the limitations and prove the correctness of our solutions. Thereby, we demonstrate how our solutions can improve reliability in TSN networks and propose these methods as guidance for users of the IEEE 802.1CB standard.
\end{abstract}


\begin{table}[b]
    \vspace{-4mm}
    \begin{tabular}{p{0.95\linewidth}}
  
  \textbf{Copyright~\copyright~2022 IEEE}\\ 
  Personal use of this material is permitted. Permission from IEEE must be obtained for all other uses, in any current or future media, including reprinting/republishing this material for advertising or promotional purposes, creating new collective works, for resale or redistribution to servers or lists, or reuse of any copyrighted component of this work in other works.\\
\textbf{Citation}:
  L. Maile, D. Voitlein, K. -S. Hielscher and R. German, "Ensuring Reliable and Predictable Behavior of IEEE 802.1CB Frame Replication and Elimination,"ICC 2022 - IEEE International Conference on Communications, Seoul, Korea, Republic of, 2022, pp. 2706-2712, doi: 10.1109/ICC45855.2022.9838905
  \\
  \textbf{Published version:} \url{https://doi.org/10.1109/ICC45855.2022.9838905}
    \end{tabular}
\end{table}

\section{Introduction}
\label{sec:intro}
With real-time communication being a key part of the fourth industrial revolution, the need for reliable industrial networks is gaining increasing importance. Time-Sensitive Networking (TSN) has been introduced to face this need. Highly time- and safety-critical applications cannot tolerate re-transmissions of lost packets. Therefore, IEEE 802.1CB-2017~\cite{8021CB} ensures ultra-reliable communication by using redundant paths to forward duplicated packets over TSN networks. 
It is the only TSN standard ensuring transmission even in the case of network failures. However, this behavior can only be achieved with a safe configuration of the network. The current standard does not offer guidance for this configuration, neither are the dangers of misconfiguration and unexpected behavior highlighted. Hofmann et al.~\cite{ChallengesAndLimitations} identified several challenges of the IEEE 802.1CB-2017 standard. However, none of these challenges have been solved yet. Thus, we propose solutions to the following two limitations identified in~\cite{ChallengesAndLimitations}:
\begin{itemize}
    \item Configuring the length of the sequence history 
    \item Studying the length of bursts in case of link failure
\end{itemize}
Besides, we reveal two additional challenges of the IEEE 802.1CB-2017 standard and propose solutions for them:
\begin{itemize}
    \item Choosing match or vector recovery algorithm
    \item Setting timer values to reset the sequence history
\end{itemize}
While some simulations have been done on the IEEE 802.1CB standard~\cite{AltFRERImp,FRERErrorSim}, to the best of our knowledge, we are the first to analyze and solve the addressed limitations. We show the correctness of our solutions both theoretically and in simulations. The proposed solutions could extend the existing standard to serve as guidance for users in future.

The IEEE 802.1CB-2017 standard is introduced in Section~\ref{sec:standard}. We describe the limitations of the standard in Section~\ref{sec:limitation}. For these limitations, we analyze the underlying effects and present our solutions in Section~\ref{sec:solution}. Section~\ref{sec:evaluation} presents the simulation results and Section~\ref{sec:conclusion} concludes the paper.

\begin{figure}[t]
\begin{center}
\includegraphics[width=\columnwidth]{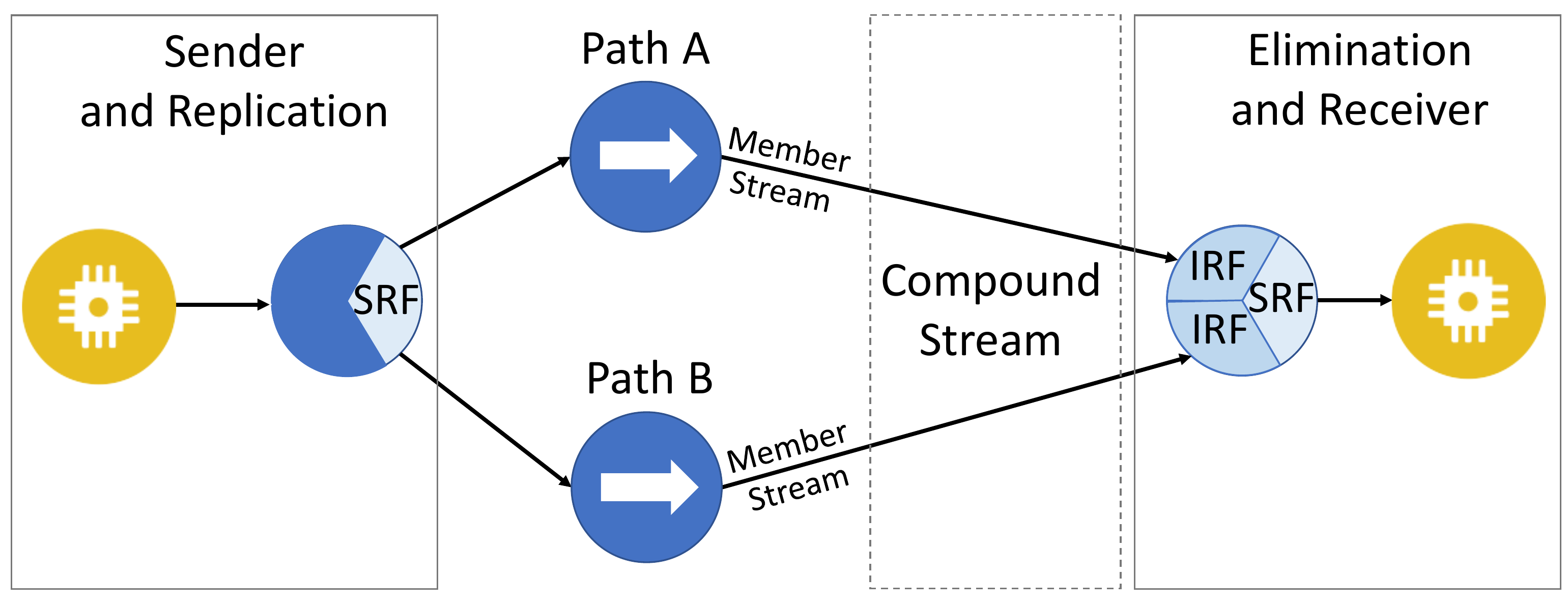}
\end{center}
\caption{ Example FRER network with two redundant paths.} 
\label{fig:network}
\vspace{-0.2cm}
\end{figure}

\section{IEEE 802.1CB-2017}
\label{sec:standard}
The full name of the TSN standard under investigation is IEEE 802.1CB-2017 Frame Replication and Elimination for Reliability, or FRER for short. As the name implies, FRER replicates packets over redundant paths and identifies and eliminates duplicate packets later in the network. FRER features replication and elimination in both end-devices and bridges, which is known as  \textit{flexible positioning}~\cite[Page~34]{8021CB}. Elimination of duplicate packets ensures that applications do not need to be aware of the redundant network operations.

Each replicated flow on a disjoint path is called a \textit{member stream}. All member streams containing the same information are grouped under the term of one \textit{compound stream}. Duplicate packets within each member stream are identified by an Individual Recovery Function (IRF). Duplicates arriving from the different member streams are identified using the Sequence Recovery Function (SRF). SRF is used to assign sequence IDs to packets before duplicating them into different member streams. Later, SRF also implements the elimination of duplicate packets. This paper focuses on SRF only. The basic components of FRER are illustrated in Fig.~\ref{fig:network}. In TSN, sending and receiving devices are respectively called talker and listener.

In TSN networks, critical streams must be registered by defining three traffic characteristics~\cite{8021Qat}: Class Measurement Interval ($\mathit{CMI}$), Max Interval Frames ($\mathit{MIF}$), and Max Frame Size ($\mathit{MFS}$). They can be interpreted as follows: A stream sends at most $\mathit{MIF}$ packets during an interval of length $\mathit{CMI}$. Each packet is smaller or equal to $\mathit{MFS}$.

The IEEE 802.1CB-2017 standard does not guarantee an in-order transmission of packets. However, since the application should not be aware of the redundancy, duplicates must be eliminated safely. Yet, it should be noted that packets arrive in FIFO order within single member streams since they are configured with static paths~\cite{ChallengesAndLimitations}.

\section{Limitations of IEEE 802.1CB-2017}
\label{sec:limitation}
IEEE 802.1CB-2017 was designed to eliminate duplicate packets and to forward all new packets, but this can only be guaranteed if the configuration is correct. Incorrectly configured networks can lead to failure of safety-relevant tasks. However, these configurations are not addressed in the standard or in other literature, leaving the decisions to the user without further information. In this section, we present four main aspects that must be considered when designing safety-critical networks using IEEE 802.1CB-2017 and for which we present the solutions in Section~\ref{sec:solution}.

\subsection{Match versus Vector Recovery Algorithm}
\label{sec:lim:MRAvsVRA}
To eliminate duplicate packets, the SRF can be configured to compare sequence numbers in one of two ways. The simplest algorithm is called Match Recovery Algorithm (MRA). It stores the highest sequence number received. Only duplicates with this sequence number are eliminated; all other packets are accepted and forwarded. By definition, MRA is only applicable to so-called intermittent streams. Intermittent streams satisfy the following requirement: when merging member streams, the difference between arrived sequence numbers cannot exceed one~\cite[Page~33]{8021CB}. If this requirement is not fulfilled, MRA is susceptible to pass duplicates and the Vector Recovery Algorithm (VRA) should be used instead.

In contrast to MRA, VRA defines an interval of sequence numbers. Within this interval, new packets are accepted and duplicates are eliminated. All packets outside this interval are discarded, even if they actually arrive for the first time. 

While MRA requires less memory and processing resources, it can only be used for the limited scenario of intermittent streams. In contrast, VRA offers more flexibility for the transmission behavior of talkers by covering a wider range of sequence numbers.

The standard does not provide guidance on how to ensure that the requirements for intermittent streams are met and, thus, MRA can be used. As far as we know, this challenge has not yet been addressed before.

\begin{figure}[t!]
\begin{center}
\includegraphics[width=1.03\columnwidth]{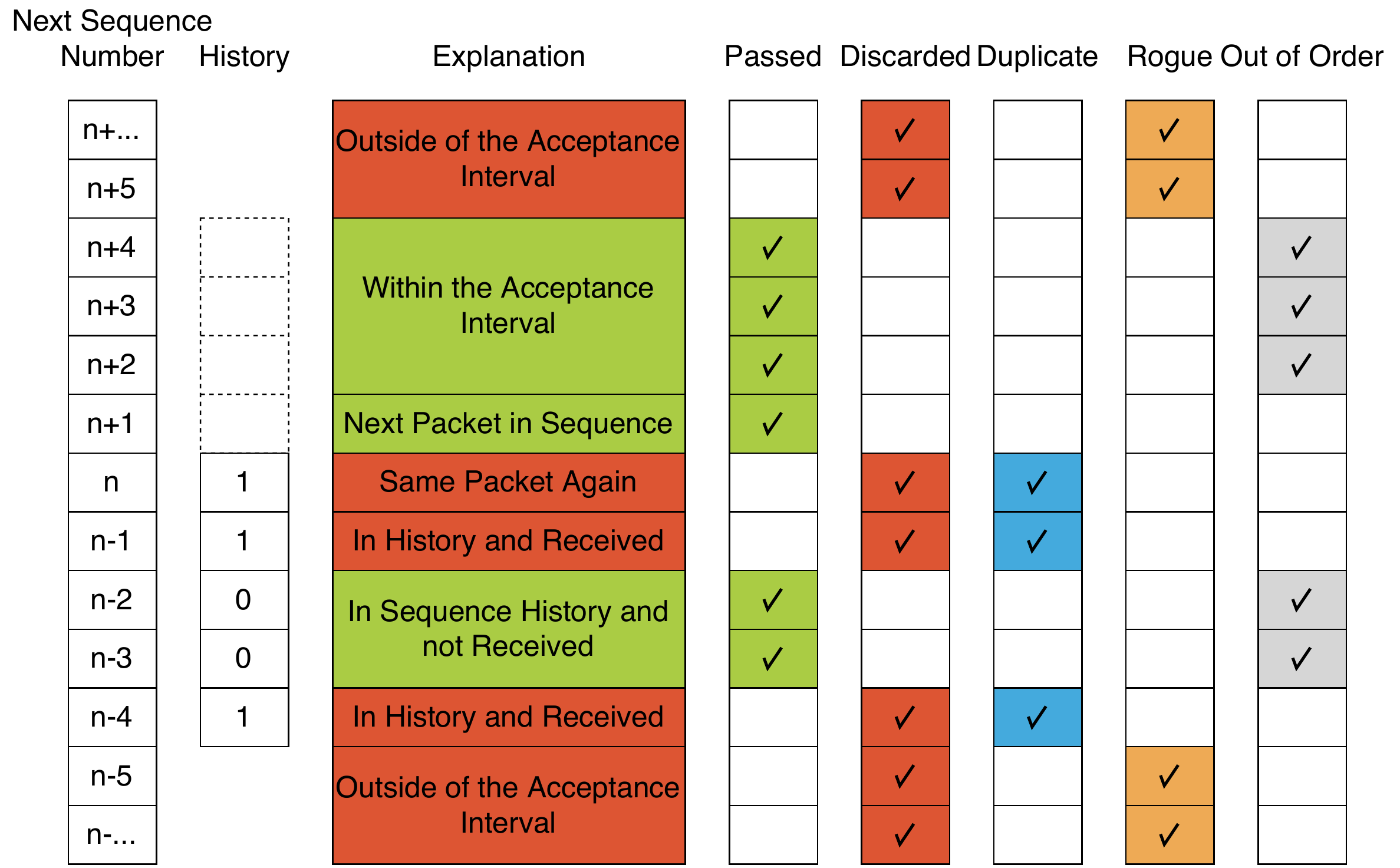} 
\end{center}
\caption{Example sequence history, illustrated in the second column. The other columns explain how a newly arriving packet would be handled, depending on its sequence number.}
\label{fig:history}
\end{figure}
 
\subsection{History Length Configuration}
\label{sec:lim:history}
If a stream requires VRA,  the user must configure the \textit{sequence number interval} for this function. This interval is defined in the standard as~\cite[Page~40]{8021CB}:
\begin{equation}
    \mathit{RecovSeqNum} \pm (\mathit{frerSeqRcvyHistoryLength} - 1)
\end{equation}
Thereby, $\mathit{RecovSeqNum}$ is initialized with the first sequence number received. Packets within the sequence number interval and higher than $\mathit{RecovSeqNum}$ lead to an update of $\mathit{RecovSeqNum}$ to the newly received sequence number. Packets that do not fall within the sequence number interval are marked as rogue and are discarded, which may cause safety-critical information to be discarded entirely. The \textit{sequence history} denotes all packets within the sequence number interval and lower than $\mathit{RecovSeqNum}$. We illustrate the handling of newly arriving packets in VRA in Fig.~\ref{fig:history}. 

The default value for $\mathit{frerSeqRcvyHistoryLength}$ in the standard is 2. However, this value must be network dependent, as also emphasized in~\cite{ChallengesAndLimitations}. 

\subsection{Reset Timer Configuration}
\label{sec:lim:reset}
IEEE 802.1CB-2017 does not offer protection against the failure of a talker or its connection to the replicating device. Instead, timers are used to respond to changes in the network. If connections are interrupted, newly arriving sequence numbers may be outside the current sequence number interval. However, packets outside the sequence number interval are discarded by VRA. Therefore, FRER triggers a reset after a period of time in which no packets have been accepted. This reset is done by the so-called $\mathit{SequenceRecoveryReset}$ function. As a result, $\mathit{RecovSeqNum}$ is reinitialized with the first arriving sequence number. The timer for the $\mathit{SequenceRecoveryReset}$ function is set in the $\mathit{frerSeqRcvyResetMSec}$ variable. Each time a new packet is passed, this timer restarts. If the timer is configured too short, duplicates will be passed because the sequence number interval has been reset. If the timer is too long, valid packets from interrupted connections will be discarded. We are the first to identify this problem and propose a solution.

\subsection{Burst Size Prediction}
\label{sec:lim:burst}
FRER is designed to always forward the first arriving packet of each sequence number to ensure low latency. However, this feature can cause packets to be delivered out of sequence and, most importantly, result in an increase of burstiness. Consider the following scenario: The delays of the redundant paths are different, and transmission errors occur on the fastest path, i.e., transmission is interrupted. This results in the following three phases: 1) Initially, only packets from the slower paths are received. However, these have already been received before via the faster path, so no new packet is forwarded. 2) Then, new packets arrive from the slower paths, which are forwarded. 3) The critical phase starts when the faster path resumes its transmission. The slower paths continue to transmit new packets. Meanwhile, packets from the faster path will arrive with sequence numbers that are higher than currently received on the slower paths. Thus, for a while, new packets are accepted from both the faster and slower paths at the same time. This results in a burst of packets as the acceptance rate doubles.

Compared to traditional burst prediction methods~\cite{chen_survey_2021} which predict future traffic characteristics in large networks, our derived solution provides an upper bound for burst lengths that are introduced solely by the usage of FRER.

Hofmann et al. state that these effects must be thoroughly investigated because "bursts of high-priority messages can have a significant impact on the timing behavior of all other frames in its path"~\cite{ChallengesAndLimitations}. Moreover, if an application buffers traffic to sort the packets, this buffer must be increased. Therefore, we add a discussion of the effects and investigate the burst dimensions in the following sections.

\section{Analysis and Solutions}
\label{sec:solution}
In this section, we present safe solutions to the above limitations. By this, we mean that no duplicate frame is passed, and no new frame is discarded. Therefore, network characteristics must be considered, in particular
\begin{itemize}
        \item best-case, i.e., lowest delay $d_{BC}$ of the fastest path and
        \item worst-case, i.e., highest delay $d_{WC}$ of the slowest path.
\end{itemize}
We refer to the derivation of these delays in Appendix~\ref{sec:appendix}. 
In the following, we refer to the difference $\Delta d = d_{WC}-d_{BC}$ for each sequence number as \textit{reception window}. For simplicity, in each section, we will first consider a periodic sending behavior with $\mathit{MIF}=1$ and afterwards present the solutions for aperiodic behavior and $\mathit{MIF}>1$. Sending devices might not send periodically, e.g., because their sending is non-deterministic and thus jitters or because the sending behavior requires aperiodicity. We refer to jitter as a talker's maximum deviation from its periodic sending time and denote it as $J$. Please note that in TSN-conformant networks $J < \mathit{CMI}$.

\subsection{Match versus Vector Recovery Algorithm}
\label{sec:sol:MRAvsVRA}
For the first limitation, we need to define whether a given compound stream is intermittent. Therefore, all copies of a packet must arrive before the next sequence number can arrive at the eliminating device~\cite[Page~33]{8021CB}. 
According to this definition, we can use MRA if and only if the $\mathit{CMI}$ for the compound stream satisfies the following equation:
\begin{equation}
    \mathit{CMI} > \Delta d = d_{WC}-d_{BC}
\end{equation}
\begin{proof}
To construct the worst case, we only need to consider the slowest and fastest paths and analyze the maximum possible difference in arriving sequence numbers. If the definition is satisfied for these paths, it is also guaranteed on all other paths. $\Delta d$ defines the total range in which all packets with one sequence number can arrive. Assuming that $\mathit{CMI} \le \Delta d$, we can construct the following scenario, also shown in Fig.~\ref{fig:overlapping}. The last duplicate of packet 101 can arrive at the end of its $\Delta d$ on the slowest path. When sending periodically each $\mathit{CMI}$, the first packet 102 may arrive earlier, violating the requirement that all duplicates must arrive before the next sequence number is received. In contrast, $\mathit{CMI} > \Delta d$ guarantees that the fastest transmission of a new packet cannot overtake the slowest transmission of a slow packet still in flight, so the requirement is met.
\end{proof}
\subsubsection{Aperiodic Traffic}
\label{sec:sol:MRAvsVRA_aper}
For aperiodic traffic, the worst case is that packet 101 is delayed by $j_1$ time units while packet 102 is sent $j_2$ time units earlier than in the periodic scenario, where $j_1 + j_2 = J \le \mathit{CMI} $. Note that if $j_1 + j_2 > \mathit{CMI}$, this would mean that packet 102 is sent before packet 101, which cannot be the case. This changes the requirement to guarantee intermittent streams to:
\begin{equation}
    \mathit{CMI} > \Delta d + J 
\end{equation}
\subsubsection{Multiple Frames}
For $\mathit{MIF}>1$, the worst case is that devices send all packets right after another, i.e., just adding interpacket gaps. This means that $\Delta d$ can be at most the duration for the reception of the shortest packet, which cannot be guaranteed in practice. Therefore, MRA can only be used for $\mathit{MIF=1}$.

\begin{figure}[t!]
\begin{center}
\includegraphics[width=1.0\columnwidth]{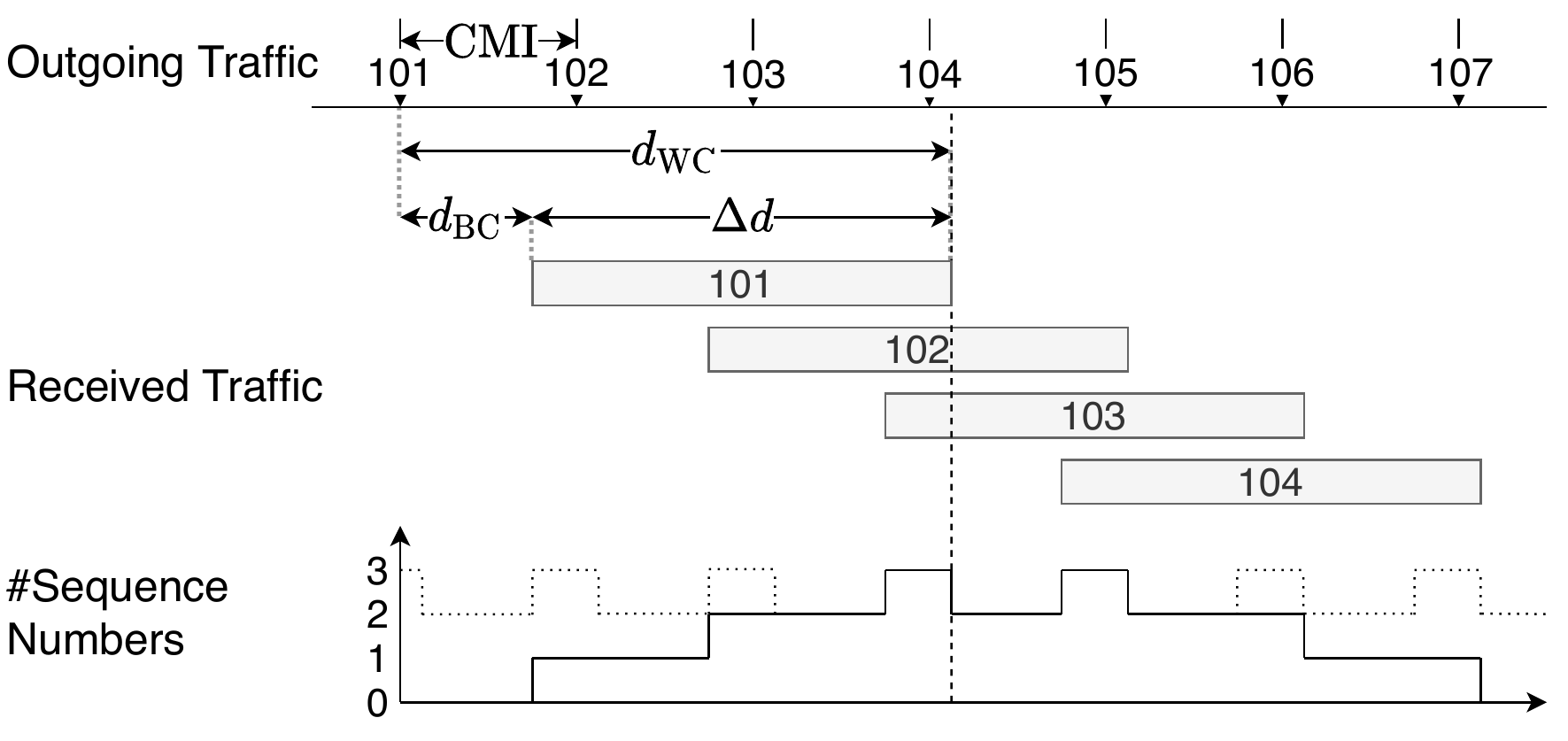} 
\end{center}
\caption{Delays and reception windows for the arrivals of packets with the same sequence number. The number of overlapping reception windows is illustrated at the bottom.} 
\label{fig:overlapping}
\end{figure}

\subsection{History Length Configuration}
\label{sec:sol:history}
To define a safe history length $L$ (short for $frerSeqRcvyHistoryLength$), we need to ensure that a sequence number is only allowed to drop out of the history if it cannot arrive anymore. Therefore, the history constantly needs an entry for each sequence number that may be received at the current time. Similar to Section~\ref{sec:sol:MRAvsVRA}, we thus need to analyze how many packets can arrive at the same time in the worst case. 
In contrast to MRA, VRA allows overlapping reception windows. Consequently, the number of overlapping reception windows is the maximum number of packets that need to be tracked at any given time. As a result, the length of the sequence history must be:
\begin{equation} \label{equ:history}
   L > \frac{\Delta d}{\mathit{CMI}} + 1
\end{equation}
\begin{proof}
For each interval of $\Delta d$, we need to determine how many other sequence numbers can overlap in the worst case. This is illustrated in Fig.~\ref{fig:overlapping}. Packets are sent in multiples of their interval $k\cdot \mathit{CMI}$. This means that any packet that is sent within a multiple of $k \le \frac{\Delta d}{\mathit{CMI}}$ represents an overlapping window. Including the sequence number for which we consider the overlapping windows, the worst integer number of overlapping sequence numbers is thus $N = \floor[\big]{\frac{\Delta d}{\mathit{CMI}}} + 1$. This results in:
\begin{equation} \label{equ:history2}
  L > \frac{\Delta d}{\mathit{CMI}}  \ge \floor[\bigg]{\frac{\Delta d}{\mathit{CMI}}} + 1 
\end{equation}
This equation holds if the sequence history always covers the range of sequence numbers that could arrive at the current time. However, only the reception of new packets triggers a shift of the sequence history. The highest sequence number that could follow directly after the current $\mathit{RecovSeqNum}$ (e.g., 101 in Fig.~\ref{fig:overlapping}) is $N$ higher (e.g., 104 in Fig.~\ref{fig:overlapping}) since it has to overlap with $\mathit{RecovSeqNum} + 1$. Otherwise, $\mathit{RecovSeqNum} + 1$ (102) is guaranteed to show up earlier. This, together with the definition of the sequence number interval, leads to~\eqref{equ:history}.
Note that higher values for the length are also safe to use, but do not provide any additional benefits. Instead, they lead to increased usage of memory resources.
\end{proof}

\subsubsection{Aperiodic Traffic}
Following the argumentation of Section~\ref{sec:sol:MRAvsVRA_aper}, we add the jitter to  $\Delta d$ which results in:
\begin{equation}
    L > \frac{\Delta d + J}{\mathit{CMI}} + 1
\end{equation}
with $J < \mathit{CMI}$, this can be simplified to:
\begin{equation}
    L > \frac{\Delta d }{\mathit{CMI}} + 2
\end{equation}

\subsubsection{Multiple Frames}
In the worst case, a stream sends one of its frames at the beginning of its interval and the remaining $(\mathit{MIF} - 1)$ packets at the end, e.g., assume multiple packets at the end of the interval for packet 101 in Fig.~\ref{fig:overlapping}. For overlapping of the following windows, not one but $\mathit{MIF}$ packets need to be considered. This maximizes the overlapping frames. Based on~\eqref{equ:history2} which again needs to be increased by one, this results in:
\begin{equation}
 L \ge  \mathit{MIF}\cdot \floor[\bigg]{\frac{\Delta d}{\mathit{CMI}} + 1 } +1+(\mathit{MIF}-1)
 =  \mathit{MIF}\cdot \floor[\bigg]{\frac{\Delta d}{\mathit{CMI}} + 2 }
\end{equation}

\subsection{Reset Timer Configuration}
\label{sec:sol:reset}

As described in Section~\ref{sec:lim:reset}, FRER uses a timeout after R~\SI{}{\milli\second} (short for $\mathit{frerSeqRcvyResetMSec}$) to react to interrupted connections and, thus, new sequence numbers. As explained in Section~\ref{sec:lim:reset}, we need to avoid both too high and two low values for this timeout. The optimal value to trigger the reset of the sequence history is:
\begin{equation} \label{equ:reset}
    R = \Delta d + \mathit{CMI}
\end{equation}
\begin{proof}
We want to trigger the reset when the sequence numbers change, e.g., because a talker loses its connection. It is safe to trigger the reset timeout when no more duplicate packets can arrive. In the worst case, exactly one packet (e.g., sequence number 101) arrives via the fastest path at the beginning of its $\Delta d$. If we reset too early, a second packet numbered 101 might show up and the sequence history will be reinitialized with that value, thereby passing a duplicate. If we reset after no packet of 101 can arrive, the sequence history will be safely reset to the newly arriving sequence number. Assuming this worst case, we see that we can safely reset the timer after $\Delta d$ time units of no packets arriving, resulting in:
\begin{equation}
    R > \Delta d
\end{equation}
This value can be applied, regardless of periodic or aperiodic traffic. However, if $\Delta d$ is small and the windows do not overlap, this configuration may result in many unnecessary resets. In this case, the reset is triggered between each sequence number, which is safe, but not required. 
This can be improved by waiting not only until no more duplicates can arrive, but until we are sure that all packets of a sequence number are lost. With periodic traffic, packets arrive after $\mathit{CMI}$ time units, resulting in~\eqref{equ:reset}.
\end{proof}

\subsubsection{Aperiodic Traffic} 
In the case of aperiodicity or jitter, the packet may arrive after $\mathit{CMI} + J$, leading to:
\begin{equation}
    R = \Delta d + J + \mathit{CMI} \text{~or~} R = \Delta d + 2\cdot \mathit{CMI}
\end{equation}

\subsubsection{Multiple Frames}
Since the timeout does not track the sequence numbers but simply the time of arrival, the same reset timer can also be used with $\mathit{MIF}>1$.

\subsection{Burst Size Prediction}
\label{sec:sol:burst}
Let us denote the maximum number of packets arriving in a burst after link failure as $n_{max}$. The maximum burst size can be calculated with:
\begin{equation} \label{equ:burst}
    n_{max} = \max \big{(}2 \cdot \ceil[\bigg]{ \frac{\Delta d}{\mathit{CMI}}} -1,\text{~}0\big{)}
\end{equation}
\begin{proof}
The maximum burst occurs when all but the slowest path fail and the fastest path is restored first. For all other path combinations, the burst size is less. If faster paths than the failed path are still operating, no burst occurs at all. The burst starts as soon as the fastest failed link recovers and the first packet arrives on that link (phase 3 in Section~\ref{sec:lim:burst}). We denote the first sequence number of the link after its recovery with $N$. The burst then lasts until $N$ arrives over the slower link that has not failed. Consequently, the duration of the burst is $\Delta d$. The maximum number of packets that can arrive during $\Delta d$ is $\ceil[\big]{ \frac{\Delta d}{\mathit{CMI}}}$~\cite{RTCal}. Since both the recovered and the slower link deliver new packets, the number of packets in the burst is twice this value. The last packet is not part of the burst because its successor is received from the same link. This results in~\eqref{equ:burst}.
\end{proof}

\begin{figure*}[!t]
  \centering
  \subfloat[Number of duplicates using MRA.]{\includegraphics[width=0.3333\textwidth]{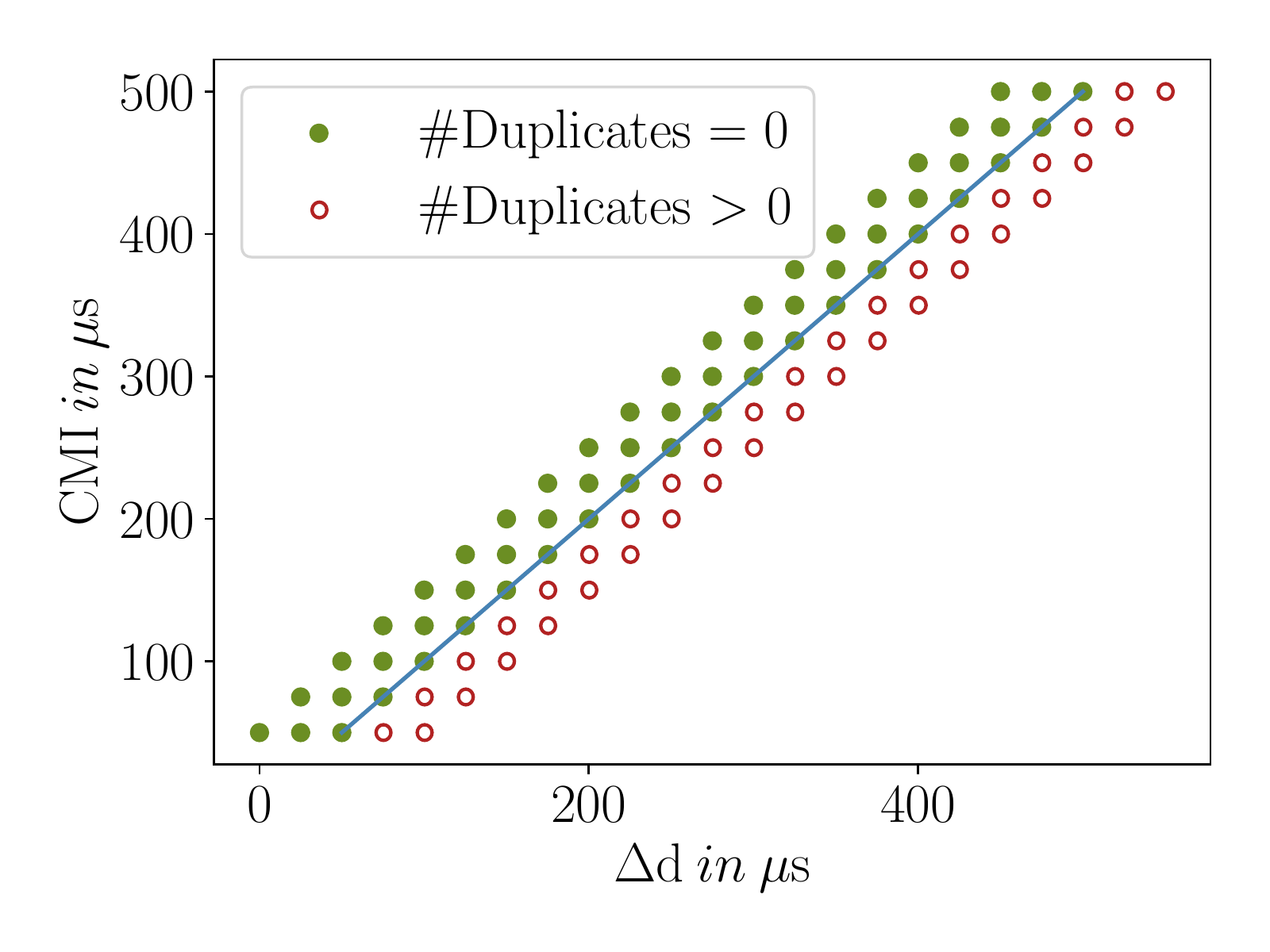}\label{fig:Match Algorithm Equation}}
  \hfill
  \subfloat[Number of rogue packets using VRA.]{\includegraphics[width=0.3333\textwidth]{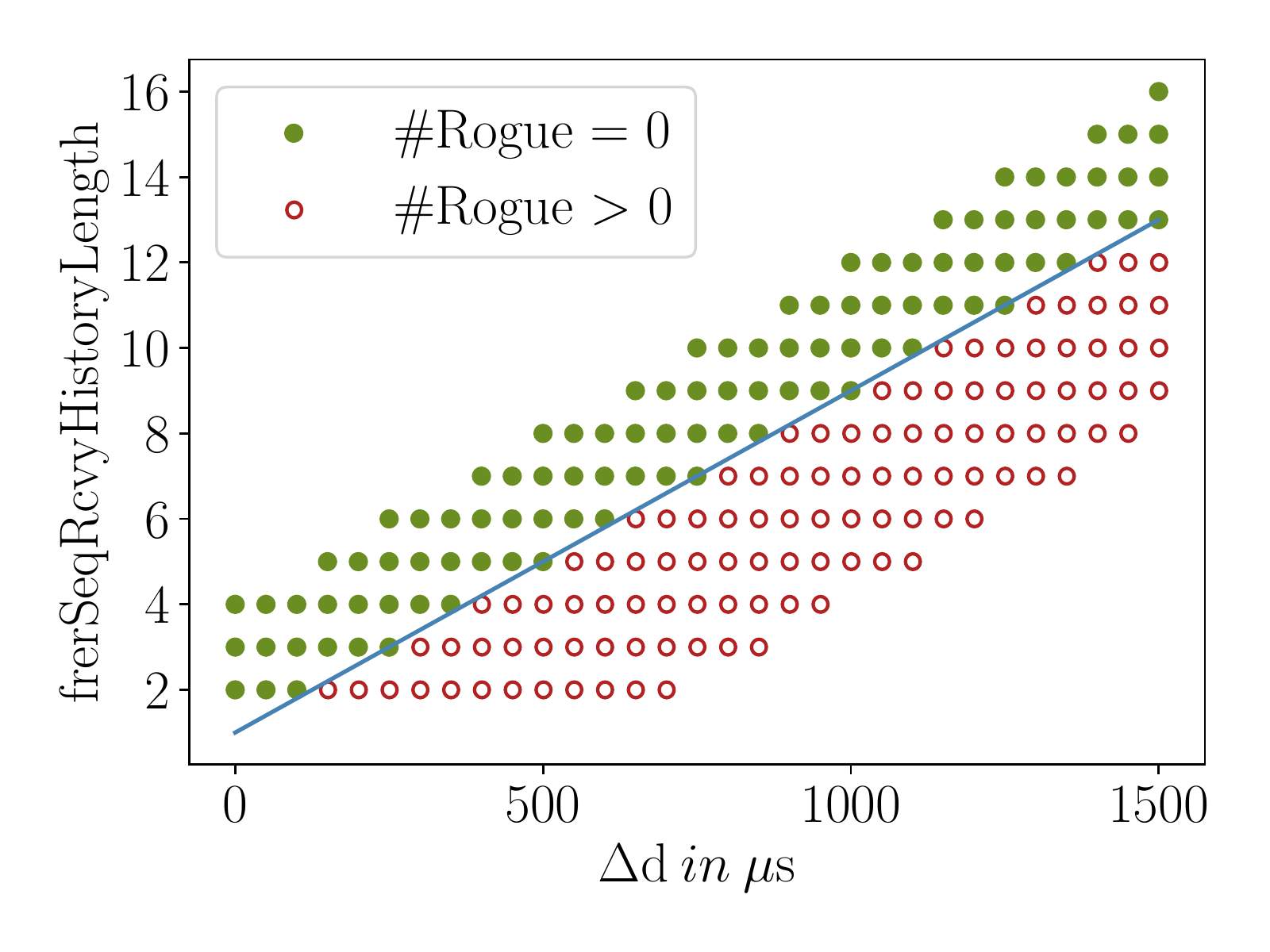}\label{fig:Vector Example}}
    \hfill
  \subfloat[Number of duplicates using VRA.]{\includegraphics[width=0.3333\textwidth]{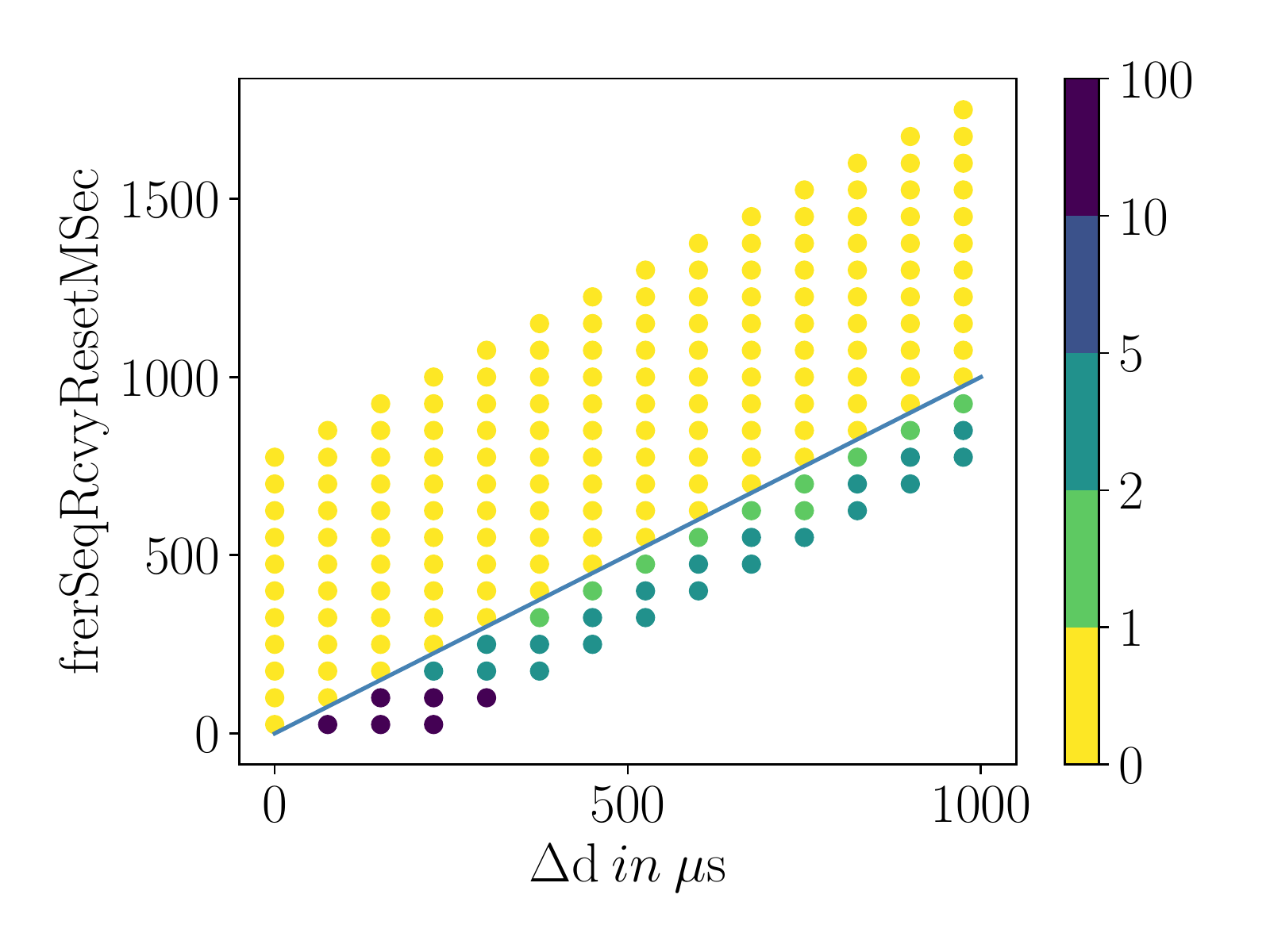}\label{fig:Reset Duplicates}}
  \hfill
  \subfloat[Number of duplicates, correctly passed packets, and resets for $\Delta d = \SI{75}{\mu\second}$.]{\includegraphics[width=0.3333\textwidth]{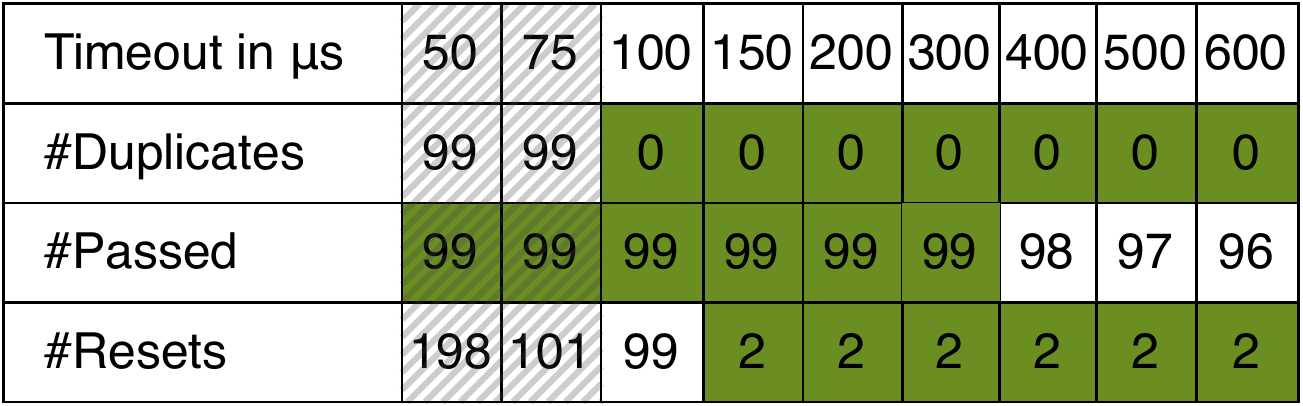}\label{fig:Resets 75}}
  \hfill
  \subfloat[Calculated versus actual burst length.]{\includegraphics[width=0.312\textwidth]{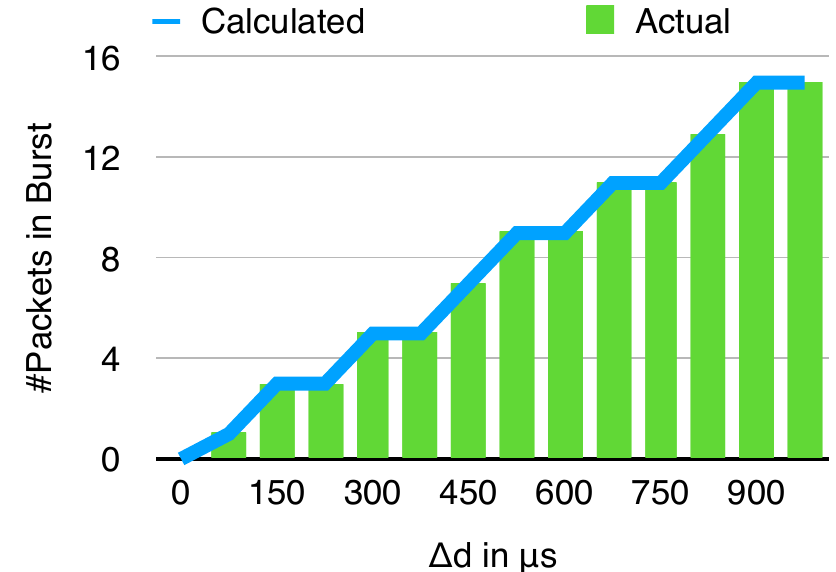}\label{fig:Burst Length}}
    \hfill
  \subfloat[TSN network with interfering traffic.]{\includegraphics[width=0.33\textwidth]{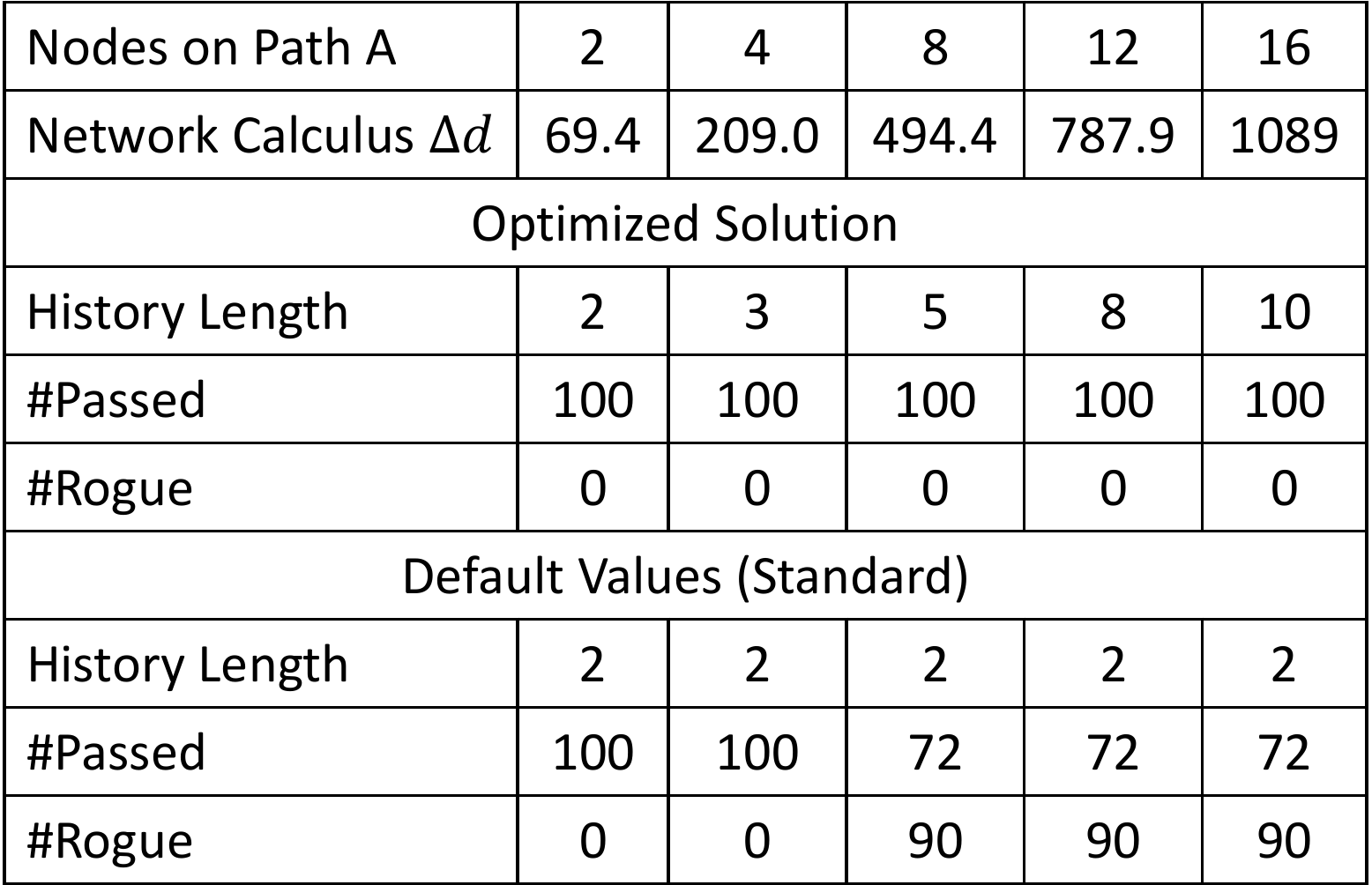}\label{fig:real example}}
  \caption{OMNeT++ simulation results for varying configurations.}
\end{figure*}

\subsubsection{Aperiodic Traffic}
As before, jitter increases the delay difference between links, resulting in a burst size of:
\begin{equation}
    n_{max} = \max \big{(2} \cdot \ceil[\bigg]{\frac{\Delta d + J}{\mathit{CMI}}} -1,\text{~}0\big{)}
\end{equation}

\subsubsection{Multiple Frames}
The worst case can be easily constructed if all $\mathit{MIF}$ packets arrive within $\Delta d$. Then, the burst is increased by the number of frames:
\begin{equation}
    n_{max} = \max \big{(}2\cdot \mathit{MIF} \cdot \ceil[\bigg]{ \frac{\Delta d}{\mathit{CMI}}} -1,\text{~}0\big{)}
\end{equation}
\vspace{+0.08cm}
\section{Evaluation}
\label{sec:evaluation}
\vspace{+0.08cm}
We used OMNeT++~\cite{omnetdoc} and the NeSTiNg library~\cite{nesting_2019} to verify the results of the equations above. The network is illustrated in Fig.~\ref{fig:network} with different path delays. For our first evaluations, the delay of each path is constant because no other traffic is present. We use the processing delays of the switches to define $\Delta d$. Unless stated otherwise, $\mathit{CMI}$ is equal to \SI{125}{\mu\second} and the talker sends 100 packets. For the history length and burst length evaluation, we introduce a \SI{75}{\milli\second} interruption on the faster path to get the worst-case behavior. For the reset timeout simulation, the connection between the talker and the duplicating device loses one packet.

\subsection{Match versus Vector Recovery Algorithm}
Fig.~\ref{fig:Match Algorithm Equation} shows the results for simulations with different $\mathit{CMI}$ and $\Delta d$ values using the MRA algorithm. We determined that a stream is intermittent when $\mathit{CMI} > \Delta d$, which corresponds to the area above the blue line in Fig.~\ref{fig:Match Algorithm Equation}. Every configuration of $\mathit{CMI}$ and $\Delta d$ that fulfilled our requirement for intermittent streams did not pass duplicates in the simulation. Also, all configurations below the line accepted at least one duplicate packet, illustrating that our boundary is not too cautious.

\subsection{History Length Configuration}
All sequence numbers arriving at the VRA must be in the sequence number interval. 
Every packet whose sequence number is not within that range increments the counter for rogue packets in eliminating devices. Fig.~\ref{fig:Vector Example} shows the number of rogue packets for a varying range of history lengths and different $\Delta d$'s. The blue line illustrates the resulting lower bound proposed by our solution. 
All optimized configurations worked in the simulations, meaning no packet was marked as rogue, i.e., arrived outside the sequence number interval. Again, the boundary given by our solution matches the boundary derived in our simulations, meaning that we do not overestimate the required history length.

\subsection{Reset Timer Configuration}
Duplicates must not pass, even when a timeout resets the recovery function. Our simulations observe the number of passed duplicates using VRA for different reset timeout values and $\Delta d$. As minimum requirement, we derived that the timeout has to be longer than $\Delta d$. In Fig.~\ref{fig:Reset Duplicates} these are the reset times above the blue line. All simulated configurations above this value successfully prevent the passing of duplicates.

As discussed in Section~\ref{sec:lim:reset}, a small $\Delta d$ can result in too many resets. Fig.~\ref{fig:Resets 75} demonstrates this effect by setting $\Delta d = \SI{75}{\mu\second}$. The timeout value for $\mathit{frerSeqRcvyResetMSec}$ is illustrated in the first line of Fig.~\ref{fig:Resets 75}. We lose one packet before its duplication to trigger the reset function. The green cells highlight optimal values for each measurement. These are: zero passed duplicates, 99 new packets - one packet is lost before the duplicating function -, and two resets - one for the lost packet and one when the talker stops sending. 

Using a timeout ${\mathit{frerSeqRcvyResetMSec}~~\le~~\Delta d}$ predicatively leads to passed duplicates. By setting $\mathit{frerSeqRcvyResetMSec} > \Delta d$, we eliminated all duplicates. However, for $\mathit{frerSeqRcvyResetMSec} \le \Delta d + \mathit{CMI} $, the number of resets can be higher than necessary. In contrast, our solution in~\eqref{equ:reset} leads to a safe configuration. As explained, values significantly higher than derived in ~\eqref{equ:reset} introduce the risk of discarding new packets. This can be seen, e.g., for timeouts values above \SI{400}{\mu\second}.

\subsection{Burst Prediction}
To determine the burst length, we count the number of packets with less than $\mathit{CMI}$ spacing to their predecessor. The bars in Fig.~\ref{fig:Burst Length} show the measured burst length for different $\Delta d$'s. The blue line is the calculated burst length in case of link failures. After the \SI{75}{\milli\second} interruption of the fastest link, the observed bursts in all simulations match the calculated lengths.

\subsection{Realistic Network}
To show the effect of our configuration in a realistic scenario, we added realistic traffic parameters and additional lower priority traffic. We use the same network structure as before, but now the switches have a constant processing delay of \SI{50}{\mu\second}~\cite{ITFLL}. The number of switches on path A can be configured dynamically to get different $\Delta d$'s. Path B uses two switches.
The definition of traffic is adapted from~\cite{matheus2021automotive, AVBBandwidth2} by using two additional traffic sources (one for each path). These sources send 1064~Bytes every \SI{100}{\mu\second}, which results in $8.6\%$ utilization. The switches use credit-based shaping with an idle slope of 0.5 for the real-time traffic and strict priority for the additional traffic. To derive the worst-case path delays, we applied Network Calculus (NC)~\cite{TSNNC}. 

Fig.~\ref{fig:real example} shows the behavior using our solutions compared to the default values in the standard. Thereby, we calculated the $\Delta d$'s and history lengths for the VRA. Each column illustrates the difference in the path delays $\Delta d$ calculated by NC in $\SI{}{\mu\second}$. We listed the results with our optimized history length in comparison to the default history length of $2$. The optimal result for passed and rogue packets is 100 and 0 respectively. As we can see in Fig.~\ref{fig:real example}, our optimized configuration safely handled all packets, whereas using the default history length results in rogue packets and packet loss with increasing path length differences.

\section{Conclusion}
\label{sec:conclusion}
We have highlighted and presented four critical limitations of the IEEE 802.1CB-2017 TSN standard. For these, we have derived safe configuration solutions for the key parameters of the sequence recovery function. We have also analyzed the maximum burst size in case of link failure. Our simulations show that the proposed solutions accurately predict whether a configuration is valid, proving the safety of our results.

In future work, we plan to show how a combined configuration approach can be implemented. This needs to include the derivation of path delays and the configuration of the network. In general, further research on worst-case and best-case delays in TSN networks is needed.

In summary, we have shown that it is possible to compute valid and safe configurations for networks using FRER. The configurations we have proposed are critical to ensure the reliability that the IEEE 802.1CB-2017 standard seeks to add. Therefore, we hope that these considerations will be helpful in future standardization processes.

\section{Acknowledgment}
The authors would like to thank Daniela Schmidt for implementing IEEE 802.1CB-2017 in the NeSTiNg library, which we could extend with our improvements.

\bibliographystyle{IEEEtran}
\bibliography{IEEEabrv,conference_101719}

\appendices
\section{Delay Analysis in TSN Networks}
\label{sec:appendix}

Safe configurations of the IEEE 802.1CB-2017 standard require the knowledge of best- and worst-case path delays. While estimations are widely used, we need safe lower and upper limits for delays to ensure safety guarantees.

\subsection{Best-Case Delay}
Considering zero delay is always a safe lower bound for the best-case scenario. Yet, this lower bound can be improved by considering hardware delays, namely propagation, transmission, and switching delays. Estimations for these delays can be found, e.g., in~\cite{ITFLL}.
\subsection{Worst-Case Delay}

Determining upper bounds on worst-case delays in TSN networks is a highly discussed research topic. In addition to hardware delays, we need to consider queuing delays due to interference of cross-traffic. Guarantees for these delays can be derived with formal analysis methods like Network Calculus~\cite{TSNNC}. Estimations have been derived, e.g., by simulations~\cite{AltFRERImp, FRERErrorSim}. When used in combination with an admission control scheme, upper bounds on arriving traffic and, thus, reservation independent worst-case delay values can be derived. This offers the possibility to minimize the necessity of reconfiguration for FRER streams. A decentral admission control scheme of TSN is introduced in~\cite{8021Qat}, while Guck et al.~\cite{Guck} propose a general solution for constant worst-case delays in centrally configured network.

\end{document}